\documentclass{elsart}

\usepackage{graphicx}

\usepackage{amssymb}

\begin{document}

\begin{frontmatter}


\title{Memory Function versus Binary Correlator in Additive Markov Chains}
\author{F.~M.~Izrailev\corauthref{cor1}}
\ead{izrailev@sirio.ifuap.buap.mx}
\corauth[cor1]{F.M.Izrailev}
\address{Instituto de F\'{\i}sica, Universidad Aut\'{o}noma de Puebla,
             Apartado Postal J-48, Puebla, Pue., 72570, M\'{e}xico}
\author{A.~A.~Krokhin}
\address{Department of Physics, University of North Texas,
             P.O. Box 311427, Denton, TX 76203}
\author{N.~M.~Makarov}
\address{Instituto de Ciencias, Universidad Aut\'{o}noma
             de Puebla, Priv. 17 Norte No. 3417, Col. San Miguel
             Hueyotlipan, Puebla, Pue., 72050, M\'{e}xico}
\author{S. S. Melnyk, O.~V.~Usatenko, V. A. Yampol'skii}
\address{A. Ya. Usikov Institute for Radiophysics and
             Electronics, Ukrainian Academy of Science,
             12 Proskura Street, 61085 Kharkov, Ukraine}




\begin{abstract}
We study properties of the additive binary Markov chain with short
and long-range correlations. A new approach is suggested that
allows one to express global statistical properties of a binary
chain in terms of the so-called memory function. The latter is
directly connected with the pair correlator of a chain via the
integral equation that is analyzed in great detail. To elucidate
the relation between the memory function and pair correlator, some
specific cases were considered that may have important
applications in different fields.
\end{abstract}

\end{frontmatter}


\section{Introduction}
\label{intro}
Much attention has been recently paid to the anomalous transport
in 1D systems with correlated disorder. In contrast with the well
studied case of a white-noise potential, the problem of
propagation of either electron or electromagnetic waves through
samples with random potentials with short or long-range
correlations, is a big theoretical challenge.

The fundamental significance of this problem is due to recent
exciting findings revising a commonly accepted belief that any
randomness in 1D structures results in the Anderson localization.
In particular, it was shown \cite{IzKr99,IzKrUll01,Lira,JPA} that
specific long-range correlations in 1D random potentials give rise
to an emergence of frequency windows of a very high and very low
transparency of waves through such potentials. It was also found
that the positions and widths of these windows can be controlled
by the form of the binary correlator of a scattering potential. A
quite simple method was suggested for constructing random
potentials that result in any predefined frequency window of an
almost perfect transparency.

From the experimental viewpoint, many of the results may have a
strong impact for the creation of a new class of electron
nanodevices, optic fibers, acoustic and electromagnetic waveguides
with selective transport properties. The predictions of the theory
have been verified experimentally \cite{KIKS00,KIKS02} by studying
transport properties of a single-mode electromagnetic waveguide
with point-like scatterers. The agreement between experimental
results and theoretical predictions was found to be unexpectedly
good. In particular, the predicted windows of a complete
reflection alternated by those of a good transparency, were
clearly observed, in spite of strong experimental imperfections.

A further development of the theory of correlated disorder is due
to the application to the surface scattering of electromagnetic
waves (or electrons). The problem of a wave propagation (both
classical and quantum) through guiding systems with corrugated
surfaces has long history and still remains a hot topic in the
literature. For single-mode waveguides with the surface scattering
the problem is equivalent to the 1D bulk scattering. For this
reason, the methods and results obtained for the latter case can
be directly applied for the waveguides
\cite{IzMak01,IzMak02PIERS,IzMak03}. Specifically it was shown,
both analytically and by direct numerical simulations, that
single-mode waveguides with a desired selective transport can be
fabricated by a proper construction of long-range-correlated
random surfaces.

The key ingredient of the theory of correlated disorder is the
binary correlator of a random potential. As was shown, for a weak
potential this correlator fully determines the
transmission/reflection of waves. The algorithm proposed in
Refs.~\cite{IzKr99,IzKrUll01,JPA} generates a statistical ensemble
of random potentials having the same binary correlator. Note that
in this method the values of the potentials are determined by a
continuous distribution function. On the other hand, in some
applications it is more appropriate to generate a correlated
potential from a finite number of elements. An important example
of such potentials is a sequence of nucleotides in a DNA molecule,
where only four elements determine a potential. In this paper we
address the mathematical problem of the construction of a
dichotomous sequence with a prescribed type of correlations. A
priory, it is not clear whether the random sequence of only two
elements may have a given binary correlator. To shed light on this
problem we concentrate our attention on statistical properties of
a binary Markov chain and their relation to the binary correlator.
Our main interest in this study is the so-called memory function
that was recently analyzed in other applications (see Refs.
[10-11] and references therein). We hope that our results may help
to extend the theory of correlated disorder to binary potentials.

The structure of this paper is as follows. In next Section 2 we
introduce an additive binary Markov chain via the memory function.
We show that the memory function can be related to the pair
(binary) correlator of a chain. In Section 3 the integral equation
for the binary correlator is developed and discussed in details.
Complimentary, in Section 4 the equation for the memory function
is derived in its general form. Next Section 5 is devoted to few
other forms of the integral equations for the memory function,
that can be more suitable in specific cases. As the simplest case,
in Section 6 we consider the Markov chain with a white-noise.
Then, in Section 7 a bit more complicated situation is discussed,
namely, the binary chain with short-range correlations of a
general type. Another interesting case of an exponentially
decreasing binary correlator is the subject of Section 8.
Unexpectedly, in this case the memory function has a very simple,
however, a quite specific form. Another modification of an
exponential pair correlator, namely, modulated by the
cosine-function, is analyzed in Section 9. In next Section 10, we
discuss a general case for which the Fourier transform of the pair
correlator has single pole in the complex plane of its argument.
Finally, in Section 11 we concentrate our attention on a very
important and quite complicated case of the step-wise power
spectrum. This case is of great importance in view of possible
applications of our theory to the transport in physical or DNA
systems described by the binary potentials.

\section{Additive Binary Markov Chain}
\label{sec-MarProc}

In what follows, we consider a \emph{homogeneous random binary
sequence} of symbols,
\begin{equation}\label{BinSeq}
\varepsilon(n)=\{\varepsilon_{0},\varepsilon_{1}\},\qquad
n\in\textbf{\textbf{Z}} =...,-2,-1,0,1,2,...
\end{equation}
To specify an $N$-\textit{step Markov chain} we introduce the
\emph{conditional probability} function,
\begin{equation}\label{ProbFunc}
P(\varepsilon(n)=\varepsilon_{0,1}\mid\varepsilon(n-1),
\varepsilon(n-2),...,\varepsilon(n-N+1),\varepsilon(n-N)).
\end{equation}
It is a probability of appearance of one of two symbols,
$\varepsilon(n)=\varepsilon_{0}$ or
$\varepsilon(n)=\varepsilon_{1}$, after a given sequence $T_{N,n}$
(''word") of length $N$,
\begin{equation}\label{TNn}
T_{N,n}=\varepsilon(n-1),\varepsilon(n-2),\dots,
\varepsilon(n-N+1),\varepsilon(n-N).
\end{equation}

The \emph{additive} Markov chain is defined by the conditional
probability function of the form,
\begin{equation}\label{AddProbFunc}
P(\varepsilon(n)=\varepsilon_{0,1}\mid T_{N,n})=
p_{0,1}+\sum_{r=1}^{N}F(r)
\frac{\varepsilon(n-r)-\overline{\varepsilon(n)}}
{\varepsilon_{0,1}-\varepsilon_{1,0}}.
\end{equation}
Here $p_{0}$ and $p_{1}$ are the parameters that imply,
respectively, the probability of occurring symbol
$\varepsilon_{0}$ and $\varepsilon_{1}$ in the whole sequence.
Evidently, $p_0+p_1=1$. For a stationary random sequence these
values $p_{0}$ and $p_{1}$ are, in fact, the only values of the
probability density of symbols $\varepsilon(n)$. Note that the
density does not depend on $n$, due to statistical homogeneity.
Here and below the bar stands for the average along a sequence. In
particular,
\begin{equation}\label{epsilon-av}
\overline{\varepsilon(n)}=\lim_{M\to\infty}\frac{1}{2M+1}
\sum_{n=-M}^{M}\varepsilon(n)=\varepsilon_0p_0+\varepsilon_1p_1.
\end{equation}
One should stress that in our case this average is equivalent to
the ensemble average.

The function $F(r)$ describes the effect of correlations between
the $n-$th symbol $\varepsilon(n)$ and $N$ previous symbols
$\varepsilon(n-r)$ with $r=1,\dots,N$. In the following we refer
to $F(r)$ as to the \emph{memory function}, since it may be
associated with a memory about statistical properties of a
sequence, necessary to generate random sequences with given
characteristics. Due to the evident condition $0\leq P(.|.)\leq1$,
this function obeys the restriction,
\begin{equation}\label{F-restr}
\sum_{r=1}^{N}|F(r)|\leq
\texttt{min}(p_0,p_1)/\texttt{max}(p_0,p_1).
\end{equation}

Our main interest is in the relevance of the memory function
$F(r)$ to such a statistical characteristic of random sequences as
the \emph{binary correlation function},
\begin{equation}\label{BinCor-def}
C(r)=\overline{\varepsilon(n)\varepsilon(n+r)}-
\overline{\varepsilon(n)}^{\,2}.
\end{equation}
The statistical homogeneity of the random sequence
$\varepsilon(n)$ provides the correlator $C(r)$ be dependent only
on the distance $r$ between two points $n$ and $n+r$. Moreover, by
definition (\ref{BinCor-def}) $C(r)$ is an even function of this
distance, $C(-r)=C(r)$. From Eqs.~(\ref{BinCor-def}) and
(\ref{epsilon-av}), it follows that the variance $C(0)$ reads
\begin{eqnarray}
\label{BinCor-Var} C(0)&=&\overline{\varepsilon^2(n)}-
\overline{\varepsilon(n)}^{\,2}=
(\varepsilon_{1}-\varepsilon_{0})^2\,p_0p_1,\,\,\,\,\,\,\mbox{with}\quad
p_0+p_1=1.
\end{eqnarray}

As was shown in Refs.~\cite{MUYaG-05,MUYa-05}, for additive Markov
chains the relation between the memory function $F(r)$ and
correlation function $C(r)$ has the form,
\begin{equation}\label{MF-BinCor}
C(r)=\sum_{r'=1}^{N}F(r')C(r-r'), \qquad r\geq1.
\end{equation}
This equation provides a possibility to obtain the memory function
$F(r)$ from the prescribed correlator $C(r)$ and to construct
effectively the Markov chain with the conditional probability
$P(.\mid .)$ according to Eq.~(\ref{AddProbFunc}).

For simplicity, below we use the binary correlator $K(r)$
normalized to the unity at $r=0$,
\begin{equation}\label{MF-BinCor-norm}
K(r)=C(r)/C(0).
\end{equation}
Also, we assume, without the loss of generality, that the length
of considered sequences is infinite. As a result, the starting
equation (\ref{MF-BinCor}) gets the form,
%
\begin{eqnarray}\label{MF-BC-eqini}
&&K(r)=\sum_{r'=1}^{\infty}F(r')K(r-r'),
\qquad r\geq1;\label{MF-BC-maineq}\\[6pt]
&&K(-r)=K(r),\qquad K(0)=1. \label{BC0}
\end{eqnarray}

Note that here the memory function $F(r)$ is defined for $r\geq1$,
while the correlator $K(r)$ being an even function, is determined
for all values of $r$. To avoid mathematical problems related to
this fact, we assume that the memory function $F(r)$ can be
analytically continued to negative values of $r$, namely,
$F(-r)=F(r)$. As one can see, we have also to specify the value of
$F(r)$ for $r=0$. In what follows, for simplicity we assume that
$F(0)=0$. As a result, we arrive at main equations that shall be
analyzed below,
%
\begin{eqnarray}\label{MF-BC-eq}
&&K(r)=\sum_{r'=-\infty}^{\infty}K(r-r')F(r')
-\sum_{r'=1}^{\infty}K(r+r'){F}(r'),\qquad r\geq1 ;\label{MF-BCor-1}\\[6pt]
&&K(-r)=K(r),\quad K(0)=1;\label{BC-even0}\\[6pt]
&&F(-r)=F(r),\quad F(0)=0.\label{MF-even0}
\end{eqnarray}

The above equations can be considered as the starting point for
two problems. The first one is to obtain the memory function
$F(r)$ by making use of Eq.~(\ref{MF-BC-eq}) in which the
correlator $K(r)$ is known. The second problem is the
complimentary one, namely, to find the correlator $K(r)$ for a
given memory function $F(r)$.

\section{Equation for Binary Correlator}
\label{sec-Eq-BC}

It is convenient to write the Fourier representation of the main
equations (\ref{MF-BC-eq}). For this, we introduce the Fourier
transform $\mathcal{K}(k)$ of the binary correlator $K(r)$, known
as the \emph{randomness power spectrum},
%
\begin{eqnarray}\label{FTr-K-def}\label{FTr-Kr}
K(r)=\frac{1}{\pi}\int_{0}^{\pi}dk\,\mathcal{K}(k)\cos(kr),
\end{eqnarray}
\begin{eqnarray}\label{FTr-Kk}
\mathcal{K}(k)=1+2\sum_{r=1}^{\infty}K(r)\cos(kr);
\end{eqnarray}
\begin{equation}\label{K-par-per}
\mathcal{K}(-k)=\mathcal{K}(k),\qquad
\mathcal{K}(k+2\pi)=\mathcal{K}(k).
\end{equation}

Since the correlator $K(r)$ is a real and even function of the
coordinate $r$, its Fourier transform (\ref{FTr-Kk}) is an even,
real and non-negative function of the wave number $k$. The
condition $K(r=0)=1$ results in the following normalization
relation for the power spectrum $\mathcal{K}(k)$,
\begin{equation}\label{Kk-norm}
\frac{1}{2\pi}\int_{-\pi}^{\pi}dk\,\mathcal{K}(k)=
\frac{1}{\pi}\int_{0}^{\pi}dk\,\mathcal{K}(k)=1\,.
\end{equation}

Analogously, the expressions for the Fourier transforms of the
memory function $F(r)$ have the form,
%
\begin{eqnarray}\label{FTr-Fr}
F(r)=\frac{1}{\pi}\int_{0}^{\pi}dk\,\mathcal{F}(k)\cos(kr),
\end{eqnarray}
\begin{eqnarray}\label{FTr-Fk}
\mathcal{F}(k)=2\sum_{r=1}^{\infty}F(r)\cos(kr);
\end{eqnarray}
\begin{equation}\label{F-par-per}
\mathcal{F}(-k)=\mathcal{F}(k),\qquad
\mathcal{F}(k+2\pi)=\mathcal{F}(k).
\end{equation}
Similarly, the Fourier transform $\mathcal{F}(k)$ of the memory
function $F(r)$ is also even and real function of the wave number
$k$. Due to the imposed condition $F(r=0)=0$, the normalization
condition for $\mathcal{F}(k)$ reads as,
\begin{equation}\label{Fk-norm}
\frac{1}{2\pi}\int_{-\pi}^{\pi}dk\,\mathcal{F}(k)=
\frac{1}{\pi}\int_{0}^{\pi}dk\,\mathcal{F}(k)=0.
\end{equation}

Now, we shall derive the Fourier transform of relation
(\ref{MF-BCor-1}). First, we represent this equation in the
following from,
\begin{eqnarray}\label{Fk-eq3}
&&\mathcal{K}(k)=1+
\sum_{r,r'=-\infty}^{\infty}\exp(ikr)K(r-r')F(r')
-\sum_{r=-\infty}^{\infty}K(r)F(r)\nonumber\\[6pt]
&&-2\sum_{r,r'=1}^{\infty}\cos(kr)K(r+r'){F}(r').
\end{eqnarray}
Now we substitute the memory function $F(r')$ due to Eq.
(\ref{FTr-Fr}), together with the use of Eqs.~(\ref{FTr-Kk}). In
this way we arrive at the relation,
\begin{eqnarray}\label{Fk-eq4}
\mathcal{K}(k)\mathcal{F}(k)-
2\sum_{r,r'=1}^{\infty}\cos(kr)K(r+r'){F}(r') =\mathcal{K}(k)+A-1,
\end{eqnarray}
where the constant $A$ is defined by
\begin{equation}\label{A-def}
A=\sum_{r=-\infty}^{\infty}K(r)\,F(r)=
\frac{1}{2\pi}\int_{-\pi}^{\pi}dk\,\mathcal{K}(k)\,\mathcal{F}(k).
\end{equation}

Into the second term of the left-hand side of Eq.~(\ref{Fk-eq4})
we substitute the correlator $K(r+r')$ in the form of the Fourier
integral (\ref{FTr-Kr}). As a result, we get the integral equation
for the power spectrum $\mathcal{K}(k)$ of the binary correlator,
\begin{equation}\label{FTr-BC-eq}
\big[\mathcal{F}(k)-1\big]\mathcal{K}(k)-
\frac{1}{2\pi}\int_{-\pi}^{\pi}dk'\,\Upsilon(k,k')\mathcal{K}(k')
=A-1.
\end{equation}

Here the integral kernel $\Upsilon(k,k')$ is described as follows,
%
\begin{eqnarray}
&&\Upsilon(k,k')=2\sum_{r=1}^{\infty}\cos(kr)\exp(-ik'r)
\times\sum_{r'=1}^{\infty}F(r')\exp(-ik'r')\label{Ups-1}\\[6pt]
&&=\frac{1}{2}\,Q(k,k')\frac{1}{\pi}\int_{0}^{\pi}dk''
\mathcal{F}(k'')Q(k'',k'),\,\,\,\,\,\,\,\,\,\qquad\label{Ups-2}
k'=\lim_{\epsilon\to+0}(k'-i\epsilon),\nonumber
\end{eqnarray}
and the $Q$-functions are defined below by Eq.~(\ref{Q-def1}).

Due to a complex nature of $Q$-functions, the kernel
$\Upsilon(k,k')$ is also a complex function. Then, in accordance
with the definition (\ref{Ups-1}), we have,
%
\begin{eqnarray}
\frac{1}{2\pi}\int_{-\pi}^{\pi}dk\,\Upsilon(k,k')=
\frac{1}{\pi}\int_{0}^{\pi}dk\,\Upsilon(k,k')=0,\label{Int-Ups1}\\[6pt]
\frac{1}{2\pi}\int_{-\pi}^{\pi}dk'\,\Upsilon(k,k')=
\frac{1}{\pi}\int_{0}^{\pi}dk'\,\Upsilon(k,k')=0.\label{Int-Ups2}
\end{eqnarray}

When integrating Eq.~(\ref{FTr-BC-eq}) over the wave number $k$
within the interval $(-\pi,\pi)$ or $(0,\pi)$, its integral term
vanishes due to Eq.(\ref{Int-Ups1}), while the equation itself
turns into identity $A-1\equiv A-1$. This fact implies that the
relation~(\ref{A-def}) automatically satisfies the solution of
Eq.~(\ref{FTr-BC-eq}), independently on the value of $A$.
Therefore, the constant $A$ is determined solely by the
normalization condition $K(0)=1$ or, the same, by
Eq.~(\ref{Kk-norm}) for the power spectrum $\mathcal{K}(k)$.
Another form of Eq.(\ref{FTr-BC-eq}) will be given below by
Eq.~(\ref{Fk-eqfin}).

\section{Equation for Memory Function}
\label{sec-Eq-MF}

Let us now substitute the memory function (\ref{FTr-Fr}) into the
left-hand side of Eq.~(\ref{Fk-eq4}). In this way we obtain the
integral equation for the Fourier transform $\mathcal{F}(k)$ of
the memory function,
\begin{equation}\label{FTr-MF-eq}
\mathcal{K}(k)\mathcal{F}(k)-
\frac{1}{\pi}\int_{0}^{\pi}dk'\,\Xi(k,k')
\mathcal{F}(k')=\mathcal{K}(k)+A-1.
\end{equation}
In this equation the kernel $\Xi(k,k')$ of the integral operator
is described by the expressions
%
\begin{eqnarray}
\Xi(k,k')&=&2\sum_{r,r'=1}^{\infty}\cos(kr)K(r+r')\cos(k'r')
\label{Ksi-1}\\[6pt]
&=&\frac{1}{4\pi}\int_{-\pi}^{\pi}dk''\mathcal{K}(k'')Q(k,k'')Q(k',k'')\,,
\label{Ksi-2}
\end{eqnarray}
where the $Q$-functions are defined as
%
\begin{eqnarray}
&&Q(k,k'')=-2\sum_{r=1}^{\infty}\cos(kr)\exp(-ik''r)\label{Q-def1}\\[6pt]
&&=1+i\frac{\sin k''}{\cos k-\cos k''}\,,\quad \quad k''=
\lim_{\epsilon\to+0}(k''-i\epsilon)\label{Q-def2}.
\end{eqnarray}
In spite of a complex nature of the $Q$-functions, the kernel
$\Xi(k,k')$ is real, even and symmetrical function of both
arguments $k$ and $k'$,
\begin{equation}\label{Ksi-prop}
\Xi(-k,k')=\Xi(k,-k')=\Xi(k,k')=\Xi(k',k).
\end{equation}
One can show that the mean value of the function $Q(k,k'')$
vanishes,
\begin{equation}\label{Int-Q}
\frac{1}{2\pi}\int_{-\pi}^{\pi}dk\,Q(k,k'')=
\frac{1}{\pi}\int_{0}^{\pi}dk\,Q(k,k'')=0.
\end{equation}
Indeed, integrating Eq.(\ref{Q-def2}) over $k$, one can obtain,
\begin{eqnarray}\label{Int1}
\frac{1}{2k_c}\int_{-k_c}^{k_c}\frac{\sin k''dk}{\cos k-\cos k''}=
=\frac{i\pi}{k_c}\Theta(k_c-|k''|)+
2\sum_{r=1}^{\infty}\frac{\sin(k_cr)}{k_cr}\sin(k''r)
\nonumber\\[6pt]
=\frac{i\pi}{k_c}\Theta(k_c-|k''|)+\frac{1}{k_c}
\ln\Bigg|\frac{\sin[(k_c+k'')/2]}{\sin[(k_c-k'')/2]}\Bigg|; \quad
0<k_c\leq\pi;\,\,\, |k''|\leq\pi.
\end{eqnarray}
The first expression for the integral is obtained by a direct
integration of the sum in Eq.(\ref{Q-def1}). In the derivation of
the second expression we have used the fact that the Heaviside
unit-step function $\Theta(x)$, for which $\Theta(x<0)=0$ and
$\Theta(x>0)=1$, arises due to the contribution of simple poles
$k=\pm(k''-i\epsilon)$, $\epsilon\to+0$. Clearly, this
contribution emerges only when the poles are within the
integration interval, i.e. if $|k''|<k_c$. The principal value
({\it P.V.}) of the integral in Eq.(\ref{Int1}) has the following
form,
\begin{eqnarray}\label{Int-log}
&&\textit { P.V.}\int_{0}^{k_c}\frac{\sin a\,dx}{\cos x-\cos a}
=\frac{1}{2}\textit { P.V.}\int_{0}^{k_c}
dx\Big[\cot\left(\frac{x+a}{2}\right)-
\cot\left(\frac{x-a}{2}\right)\Big]\nonumber\\[6pt]
&&= \ln\Bigg|\frac{\sin[(k_c+a)/2]}{\sin[(k_c-a)/2]}\Bigg|.
\end{eqnarray}
One can see that the integral (\ref{Int-log}) vanishes when
$k_c=\pi$, and the $Q$-function property (\ref{Int-Q}) is evident.

Note that Eq.(\ref{Int-Q}) gives rise to the normalization
relations for the kernel $\Xi(k,k')$,
%
\begin{eqnarray}
\frac{1}{2\pi}\int_{-\pi}^{\pi}dk\,\Xi(k,k')=
\frac{1}{\pi}\int_{0}^{\pi}dk\,\Xi(k,k')=0,\label{Int-Ksi1}\\[6pt]
\frac{1}{2\pi}\int_{-\pi}^{\pi}dk'\,\Xi(k,k')=
\frac{1}{\pi}\int_{0}^{\pi}dk'\,\Xi(k,k')=0.\label{Int-Ksi2}
\end{eqnarray}
As a result, the integration of Eq.~(\ref{FTr-MF-eq}) over the
wave number $k$ within the interval $(-\pi,\pi)$ or $(0,\pi)$
leads to the identity $A\equiv A$. This fact implies that the
relation~(\ref{A-def}) automatically satisfies the solution of
Eq.~(\ref{FTr-MF-eq}). Therefore, the constant $A$ is determined
by the initial condition for the memory function $F(r)$,
\begin{equation}\label{A-eq}
F(0)=\frac{1}{2\pi}\int_{-\pi}^{\pi}dk\,\mathcal{F}(k)=
\frac{1}{\pi}\int_{0}^{\pi}dk\,\mathcal{F}(k)=0.
\end{equation}

\section{Other Forms of Equation for Memory Function}
\label{sec-Eq-MF0}

In some cases, instead of Eq.~(\ref{FTr-MF-eq}) it is more
convenient to use other forms of the integral equation for the
Fourier transform $\mathcal{F}(k)$ of the memory function.
Specifically, the expression (\ref{Ksi-1}) for the integral kernel
$\Xi(k,k')$ turns out to be ineffective for some kinds of the
correlator $K(r)$. In this case one should use Eq.~(\ref{Ksi-2})
and, therefore, deal with complex $Q$-functions (\ref{Q-def1}),
taking explicitly into account their pole contributions. In what
follows, we assume that $\mathcal{K}(k)$ and $\mathcal{F}(k)$ are
\textit{even}, \textit{periodic} and \textit{sectionally
continuous} functions on real axis $k$, see Eqs.~(\ref{K-par-per})
and (\ref{F-par-per}).

In accordance with Eq.~(\ref{Q-def2}) the product of two
$Q$-functions can be written as
\begin{eqnarray}
&&Q(k,k'')Q(k',k'')=1+i\frac{\sin k''}{\cos k-\cos k''}\label{QQ}\\[6pt]
&&+i\frac{\sin k''}{\cos k'-\cos k''}-\frac{\sin^2k''}{(\cos
k-\cos k'')(\cos k'-\cos k'')}\,.\nonumber
\end{eqnarray}
After the substitution of Eq.~(\ref{QQ}) into the definition
(\ref{Ksi-2}) for the kernel $\Xi(k,k')$, one can obtain,
\begin{equation}\label{Ksi-Ksi0}
2\,\Xi(k,k')=1-\mathcal{K}(k)-\mathcal{K}(k')-2\Xi_{0}(k,k').
\end{equation}
Here the first term arises due to the unit in Eq.~(\ref{QQ}) and
normalization condition (\ref{Kk-norm}). The second and third
terms are direct consequences of the following relation,
\begin{eqnarray}\label{Int-Kpi}
\frac{1}{2\pi}\int_{-\pi}^{\pi}dk''\mathcal{K}(k'')\,\frac{\sin
k''}{\cos k-\cos k''}=i\mathcal{K}(k), \quad \quad |k|\leq\pi.
\end{eqnarray}
This relation can be obtained from a more general one,
\begin{eqnarray}\label{Int-Kkc}
\frac{1}{2k_c}\int_{-k_c}^{k_c}dk''\mathcal{K}(k'')\,\frac{\sin
k''}{\cos k-\cos k''}
=\frac{i\pi}{k_c}\Theta(k_c-|k|)\mathcal{K}(k),\,\nonumber\\[6pt]
\qquad0<k_c\leq\pi,\quad |k|\leq\pi.
\end{eqnarray}
The integral in Eq.(\ref{Int-Kkc}) is determined by simple poles
$k''-i\epsilon=\pm|k|$, $\epsilon\to+0$, if the poles are within
the integration interval, i.e. when $|k|<k_c$. The corresponding
principal value of the integral vanishes because the integrand is
an odd function. Thus, when $k_c=\pi$, Eq.~(\ref{Int-Kkc}) reduces
to Eq.~(\ref{Int-Kpi}).

One can see that in Eq.~(\ref{Ksi-Ksi0}) a new integral kernel
\begin{eqnarray}\label{Ksi0-def}
\Xi_{0}(k,k')=\frac{1}{4\pi}\int_{-\pi}^{\pi}dk''\mathcal{K}(k'')
\frac{\sin^2k''}{(\cos k-\cos k'')(\cos k'-\cos k'')}
\end{eqnarray}
is introduced. Similar to the function $\Xi(k,k')$, the new kernel
$\Xi_{0}(k,k')$ is a real, even and symmetrical function with
respect to both arguments $k$ and $k'$,
\begin{equation}\label{Ksi0-prop}
\Xi_{0}(-k,k')=\Xi_{0}(k,-k')=\Xi_{0}(k,k')=\Xi_{0}(k',k).
\end{equation}
In line with Eq.~(\ref{Int1}) taken at $k_c=\pi$ and
Eq.~(\ref{Int-Kpi}), this new kernel $\Xi_{0}(k,k')$ satisfies to
the following integral properties,
%
\begin{eqnarray}
\frac{1}{\pi}\int_{0}^{\pi}dk\,\Xi_{0}(k,k')=
-\frac{1}{2}\,\mathcal{K}(k')\,,
\label{Ksi0-int1}\\[6pt]
\frac{1}{\pi}\int_{0}^{\pi}dk'\,\Xi_{0}(k,k')
=-\frac{1}{2}\,\mathcal{K}(k)\,. \label{Ksi0-int2}
\end{eqnarray}
These relations, together with Eq.~(\ref{Ksi-Ksi0}) and
normalization condition (\ref{Kk-norm}), provide the properties
(\ref{Int-Ksi1}), (\ref{Int-Ksi2}) of the old kernel $\Xi(k,k')$.

Finally, with the use of Eq.~(\ref{Ksi-Ksi0}), the normalization
condition (\ref{A-eq}) for the memory function, and the definition
(\ref{A-def}) for the constant $A$, we arrive at the new
equivalent integral equation for the memory function in
$k$-representation,
\begin{equation}\label{Fk-eq0}
\mathcal{K}(k)\mathcal{F}(k)+
\frac{1}{\pi}\int_{0}^{\pi}dk'\,\Xi_{0}(k,k')
\mathcal{F}(k')=\mathcal{K}(k)+\frac{1}{2}A-1.
\end{equation}
As one should expect, when integrating Eq.~(\ref{Fk-eq0}) over the
wave number $k$ within the interval $(-\pi,\pi)$ or $(0,\pi)$, its
integral term gives the value $-A/2$ due to the equality
(\ref{Ksi0-int1}), thus, reducing the equation into the identity,
$A/2\equiv A/2$.

Now let us derive one more form of the integral equation. To this
end, one should substitute the expression (\ref{Ksi0-def}) into
the integral term of Eq.~(\ref{Fk-eq0}), with a further change of
the order of integration over $k$ and $k''$,
\begin{eqnarray}\label{Int-FKsi0}
&&\frac{1}{\pi}\int_{0}^{\pi}dk'\,\Xi_{0}(k,k')\mathcal{F}(k')
\nonumber\\[6pt]
&&=\frac{1}{4\pi}\int_{-\pi}^{\pi}dk''\mathcal{K}(k'')\frac{\sin
k''}{\cos k-\cos k''}
\times\frac{1}{\pi}\int_{0}^{\pi}dk'\,\mathcal{F}(k')\frac{\sin
k''}{\cos k'-\cos k''}
\nonumber\\[6pt]
&&=\frac{1}{4\pi}\int_{-\pi}^{\pi}dk''\mathcal{K}(k'')\frac{\sin
k''}{\cos k-\cos k''} \times\Big[i\mathcal{F}(k'')+\frac{1}{\pi}
\textit { P.V.} \int_{0}^{\pi}dk'\,\mathcal{F}(k')\frac{\sin
k''}{\cos k'-\cos k''}\Big]
\nonumber\\[6pt]
&&=-\frac{1}{2}\mathcal{K}(k)\mathcal{F}(k)+\frac{1}{4\pi}
\int_{-\pi}^{\pi}dk''\mathcal{K}(k'')\frac{\sin k''}{\cos k-\cos
k''}
\nonumber\\[6pt]
&&\times\frac{1}{\pi} \textit {
P.V.}\int_{0}^{\pi}dk'\,\mathcal{F}(k')\frac{\sin k''}{\cos
k'-\cos k''}\,.
\end{eqnarray}
Here we have employed the ideas used to evaluate the integral
(\ref{Int1}), and the formula (\ref{Int-Kpi}). After we substitute
the result (\ref{Int-FKsi0}) into Eq.~(\ref{Fk-eq0}), the integral
equation reads
\begin{eqnarray}
&&\mathcal{K}(k)\mathcal{F}(k)+\frac{1}{2\pi^2}
\int_{-\pi}^{\pi}dk''\mathcal{K}(k'')\frac{\sin k''}{\cos k-\cos
k''}\label{Fk-eqfin}\\[6pt]
&&\times \textit {
P.V.}\int_{0}^{\pi}dk'\,\mathcal{F}(k')\frac{\sin k''}{\cos
k'-\cos k''}=2\mathcal{K}(k)+A-2\,.\nonumber
\end{eqnarray}
As above, the symbol ``$\textit { P.V.}$'' denotes a principal
value of the corresponding integral. Since we cannot change the
integration order in Eq.~(\ref{Fk-eqfin}), its application is
suitable if the power spectrum $\mathcal{K}(k)$ is considered as
unknown while the memory function $\mathcal{F}(k)$ is predefined.
When integrating Eq.(\ref{Fk-eqfin}) over $k$ within the interval
$(-\pi,\pi)$ or $(0,\pi)$, its integral term vanishes due to
Eq.~(\ref{Int1}), while the equation itself turns into the
identity $A\equiv A$.

One should, however, emphasize that all the integral equations
(\ref{FTr-MF-eq}), (\ref{Fk-eq0}), and (\ref{Fk-eqfin}) cannot be
solved in general case of any form of the additive-Markov-chain
power spectrum $\mathcal{K}(k)$. This is because of a quite
complicated structure of the integral kernels. For this reason,
below we shall analyze some particular cases when the solution can
be obtained in a relatively simple way.

\section{White-Noise Disorder}
\label{sec-WND}

The white-noise Markov chain is specified by the binary correlator
and corresponding power spectrum as follows,
%
\begin{eqnarray}
K_{wn}(r)=\delta_{r,0}\,\,,\quad \quad \quad \quad
\mathcal{K}_{wn}(k)&=&1. \label{WN-Kr}
\end{eqnarray}
In order to obtain the memory function $F_{wn}(r)$ of the additive
chain, one can use the equation (\ref{MF-BC-maineq}), or
(\ref{MF-BCor-1}). Due to the restriction $r\geq1$, one should
substitute zero in their left-hand sides, that gives rise to the
expected result,
\begin{equation}\label{WN-F}
F_{wn}(r)=0.
\end{equation}

Now, let us see how the same result follows from the integral
equations (\ref{FTr-MF-eq}), (\ref{Fk-eq0}) and (\ref{Fk-eqfin}).
One should take into account that for a white noise with
$\mathcal{K}_{wn}(k)=1$ the definition (\ref{A-def}) for $A$ and
the normalization condition (\ref{A-eq}) yield $A=F(0)=0$.
Therefore, the right-hand side of the equations vanishes and they
become homogeneous, with a trivial solution,
\begin{equation}\label{WN-Feq}
\mathcal{F}_{wn}(k)=0.
\end{equation}
In such a way, we again come to the result (\ref{WN-F}).

Let us write down the integral kernels of the equations with the
white-noise power spectrum $\mathcal{K}_{wn}(k)=1$. First, for
Eq.~(\ref{FTr-MF-eq}) we have,
\begin{eqnarray}\label{WN-Ksi}
\Xi_{wn}(k,k')&=&2\sum_{r,r'=1}^{\infty}\cos(kr)\delta_{r+r',0}\cos(k'r')
\nonumber\\[6pt]
&=&\frac{1}{4\pi}\int_{-\pi}^{\pi}dk''Q(k,k'')Q(k',k'')=0\,.
\end{eqnarray}
In this expression the sum is evidently equal to zero. To make
sure explicitly that the integral also vanishes, one should employ
Eqs.~(\ref{QQ}), (\ref{Int-Kpi}) and the relation
\begin{equation}\label{Int4-pi}
\frac{1}{2\pi}\int_{-\pi}^{\pi}\frac{\sin^2k''dk''}{(\cos k-\cos
k'')(\cos k'-\cos k'')}=-1\,.
\end{equation}
As for Eq.~(\ref{Fk-eq0}), taking into account
Eq.~(\ref{Int4-pi}), we get
\begin{equation}\label{WN-Ksi0}
\Xi_0^{wn}(k,k')=-1/2\,.
\end{equation}

\section{Short-Range Correlations}
\label{sec-ShRCorr}

The simplest non-trivial case for which one can obtain an
asymptotically exact solution of Eq.~(\ref{FTr-MF-eq}), is the
additive Markov chain with \textit{short range correlations}. In
this case the correlator $K(r)$ is assumed to be a rapidly
decreasing function of the index $r$, with a ``very short" scale
$R_c>0$ of its decrease. It is evident that for short-range
correlations the power spectrum $\mathcal{K}(k)$ is a very smooth
function of the wave number $k$ with a ``long" scale
$k_c=R_c^{-1}>0$ of decrease. One should take into account a
discrete nature of the argument $r$ of the correlator $K(r)$, or,
the same, the periodicity (\ref{K-par-per}) of its power spectrum
$\mathcal{K}(k)$. Then, it becomes clear that depending on a
specific form of $K(r)$ (or $\mathcal{K}(k)$), the definition of
short range correlations assumes one of two conditions,
\begin{equation}\label{SRCor}
R_c=k_c^{-1}\ll1,\qquad\mbox{or}\quad \pi-k_c\ll\pi.
\end{equation}

In this case the kernel $\Xi(k,k')$ of the integral operator, see
Eq.~(\ref{Ksi-1}), is mainly contributed by values $r=r'=1$.
Consequently, one can get the following estimate for $\Xi(k,k')$,
\begin{equation}\label{SRCor-Ksi}
\Xi(k,k')\approx 2K(2)\cos k \cos k'\ll K(0)=1.
\end{equation}
Therefore, the integral term in Eq.~(\ref{FTr-MF-eq}) can be
neglected. As a result, the integral equation transforms into an
algebraic one with the solution,
\begin{equation}\label{SRCor-Fk}
\mathcal{F}(k)=1-(1-A)\mathcal{K}^{-1}(k).
\end{equation}
By making use of Eqs.~(\ref{FTr-Fr}) and (\ref{SRCor-Fk}), the
memory function in the coordinate representation reads
\begin{equation}\label{SRCor-Fr}
F(r)=\delta_{r,0}-(1-A)\,I(r)\,,
\end{equation}
with the integral $I(r)$ defined by
\begin{equation}\label{I-def}
I(r)=\frac{1}{2\pi}\int_{-\pi}^{\pi}dk\,
\frac{\exp(-ikr)}{\mathcal{K}(k)} =\frac{1}{\pi}\int_{0}^{\pi}dk\,
\frac{\cos(kr)}{\mathcal{K}(k)}.
\end{equation}

Applying the initial condition (\ref{A-eq}), one can easily get
\begin{equation}\label{SRCor-A}
1-A=I^{-1}(0).
\end{equation}

As a result, in the case of short-range correlations we arrive at
a very simple expression for the memory function $F(r)$,
\begin{equation}\label{SRCor-finFr}
F(r)=\delta_{r,0}-I(r)/I(0).
\end{equation}
Evidently, this expression should have a crossover to
Eq.~(\ref{WN-F}) for a white-noise disorder. Indeed, if
$\mathcal{K}(k)=1$, one easily gets $I(r)=\delta_{r,0}$ and
$F(r)=0$.

\section{Exponential Correlations}
\label{sec-ExpCorr}

Clearly, the equation (\ref{FTr-MF-eq}) can be easily solved if
the kernel $\Xi(k,k')$, see Eq.~(\ref{Ksi-1}), is factorized by a
product of two terms that separately depend on $k$ and $k'$ only.

A particular important example, giving rise to such a form of the
kernel $\Xi(k,k')$, is the Markov chain with an
\textit{exponential correlator},
%
\begin{eqnarray}
K_{e}(r)=\exp(-k_c|r|)\,,\quad \quad \quad
\mathcal{K}_{e}(k)&=&\frac{\sinh k_c}{\cosh k_c-\cos k}\,.
\label{Exp-Kk}
\end{eqnarray}
Here $k_c=R_c^{-1}>0$ determines the inverse correlation length.
The explicit form (\ref{Exp-Kk}) for the randomness power spectrum
$\mathcal{K}(k)$ can be found with the use of the relation,
\begin{equation}\label{Exp-GrRyzh}
2\sum_{r=1}^{\infty}\cos(kr)\exp(-k_cr)= \frac{\sinh k_c}{\cosh
k_c-\cos k}-1,
\end{equation}
that can be obtained by a direct summation of the geometric
progression, after rewriting $\cos(kr)$ in the Euler form.

In accordance with Eqs.~(\ref{Ksi-1}), (\ref{Exp-Kk}), and
(\ref{Exp-GrRyzh}), the kernel of the integral operator
$\Xi(k,k')$ can be written in the factorized form,
\begin{equation} \label{Exp-Ksi}
\Xi_e(k,k')=\frac{1}{2}\left[\mathcal{K}_{e}(k)-1\right]
\left[\mathcal{K}_{e}(k')-1\right].
\end{equation}
This form provides a possibility to find easily the solution of
Eq.~(\ref{FTr-MF-eq}),
%
\begin{equation}\label{Exp-Fk-sol}
\mathcal{F}(k)=\frac{2I(0)}{1+I(0)}\left[1-1/I(0)\mathcal{K}_{e}(k)\right],
\end{equation}
\begin{equation}\label{Exp-Fr-sol}
F(r)=\frac{2I(0)}{1+I(0)}\left[\delta_{r,0}-I(r)/I(0)\right],
\end{equation}
with $I(r)$ given by the definition (\ref{I-def}).

Taking into account the expression (\ref{Exp-Kk}) for
$\mathcal{K}(k)$ when evaluating the integral in Eq.(\ref{I-def}),
one can get
\begin{equation}\label{Exp-I}
I(r)=\coth k_c\,\delta_{r,0}-(2\sinh k_c)^{-1}\delta_{|r|,1}\,.
\end{equation}
Finally, we come to the expression for the memory function,
\begin{equation}\label{Exp-Fr}
F(r)=\exp(-k_c)\delta_{|r|,1}\,.
\end{equation}

It is important that the additive binary Markov chain with an
exponential correlator is nothing but a sequence with the one-step
memory function. As one can see, the exponential nature of the
correlator $K(r)$ results in a one-step-form of the memory
function $F(r)$, while its variation scale determines the
amplitude only.

For the Markov chain of symbols
$\{\varepsilon_0=0,\varepsilon_1=1\}$ with the relative number
$p_1$ of ''1'' in the chain, the conditional probability of the
symbol ''1'' occurring after the symbol ''1'' is equal to
\begin{equation}\label{Exp-P11}
P(\varepsilon(i)=1\mid1)=p_1+\exp(-k_c)(1-p_1).
\end{equation}
Here we took into account that
$\langle\varepsilon(i)\rangle=p_0\times0+p_1\times1=p_1$. In the
limit of an anomalously small correlation length $k_c^{-1}\to 0$,
the conditional probability function tends to constant value,
$P(\varepsilon(n)=1\mid1)\to p_1$ that is independent of previous
symbols. This limit evidently corresponds to the uncorrelated
sequence. In the opposite case of long-range correlations,
$k_c^{-1}\to\infty$, the conditional probability function is
almost equal to unity, $P(\varepsilon(n)=1\mid1)\to1$. In this
case we have a sequence with a maximal persistent diffusion.

There is another important example for which the kernel of the
operator $\Xi(k,k')$ has a separable form. Namely, this is the
Markov chain with  an exponential \emph{alternating-sign
correlator},
%
\begin{eqnarray}
K(r)=(-1)^{r}\exp(-k_c|r|)\,,\quad \quad \quad
\mathcal{K}(k)=\frac{\sinh k_c}{\cosh k_c+\cos k}\,.
\label{ASExp-Kr}
\end{eqnarray}
For this case, we obtain the following integral $I(r)$,
\begin{equation}\label{AsExp-I}
I(r)=\coth k_c\,\delta_{r,0}+(2\sinh k_c)^{-1}\delta_{|r|,1}\,,
\end{equation}
and, correspondingly, the negative memory function,
\begin{eqnarray}\label{ASExp-Fr}
F(r)=\frac{2I(0)}{1+I(0)}\left[\delta_{r,0}-I(r)/I(0)\right]
=-\exp(-k_c)\delta_{|r|,1}\,.
\end{eqnarray}
As is known~\cite{MUYaG-05}, the negative memory function
describes the anti-persistent correlated diffusion in the additive
binary Markov chain.

To conclude, we would like to note that in the limit of
short-range correlations with $k_c\gg1$, one gets $I(0)\approx1$.
This leads the exact solutions (\ref{Exp-Fr}) and (\ref{ASExp-Fr})
that coincide with the asymptotic expression (\ref{SRCor-finFr}).

\section{Exponential-Cosine Correlations}
\label{sec-ExpCosCorr}

In previous Section \ref{sec-ExpCorr} we have revealed that the
exponential binary correlator (\ref{Exp-Kk}) results in the
one-step memory function (\ref{Exp-Fr}). In order to have a
many-steps memory function, we consider the oscillating correlator
with an exponential decrease of its amplitude,
%
\begin{eqnarray}
K_{ec}(r)=\exp(-k_0|r|)\cos(\omega r)
=\frac{1}{2}\,[K_e(r)+K_e^*(r)];\label{ExpCos-Kr}
\end{eqnarray}
\begin{eqnarray}
\mathcal{K}_{ec}(k)=\frac{(\cosh k_0-\cos\omega\cos k)\sinh k_0}
{(\cosh k_0\cos\omega-\cos
k)^2+\sinh^2k_0\sin^2\omega}\label{ExpCos-Kk}\\[6pt]
=\frac{1}{2}\,[\mathcal{K}_{e}(k)+\mathcal{K}_{e}^*(k)]
\label{ExpCos-Kkec-e}\,.
\end{eqnarray}
Here the asterisk ``$*$" stands for the complex conjugation. The
last expressions (\ref{ExpCos-Kr}) and (\ref{ExpCos-Kkec-e}) are
given to relate the exp-cosine correlator and its power spectrum
to those (\ref{Exp-Kk}) for purely exponential correlations. Based
on this relationship, one should perform the following complex
substitution,
\begin{equation}\label{ExpCos-kc}
k_c=k_0+i\omega.
\end{equation}

In line with the definition (\ref{Ksi-1}) and representation
(\ref{ExpCos-Kr}), the kernel $\Xi(k,k')$ of the integral operator
is described by the sum
\begin{equation}\label{ExpCos-Ksi}
\Xi_{ec}(k,k')=\frac{1}{2}\,[\Xi_e(k,k')+\Xi_e^*(k,k')].
\end{equation}
Therefore, according to the factorized form (\ref{Exp-Ksi}) of the
operator $\Xi_e(k,k')$, the integral equation (\ref{FTr-MF-eq})
gives rise to the following solution for the memory function,
%
\begin{eqnarray}
\mathcal{F}_{ec}(k)=\frac{2+A}{2}-
\frac{2-A}{2}\,\frac{1}{\mathcal{K}_{ec}(k)}
+\frac{B}{2}\,\frac{\mathcal{K}_{e}(k)-\mathcal{K}_{e}^*(k)}
{2\mathcal{K}_{ec}(k)}\label{ExpCos-Fk-sol}\,,
\end{eqnarray}
\begin{eqnarray}
F_{ec}(r)=\frac{2+A}{2}\,\delta_{r,0}-\frac{2-A}{2}\,I(r)
+\frac{B}{2}\,\Phi(r)\,.\label{ExpCos-Fr-sol}
\end{eqnarray}
Here the real constant $A$ is defined by the standard expression
(\ref{A-def}), while the imaginary constant $B$ is defined by
\begin{equation}\label{ExpCos-B-def}
B=\frac{1}{\pi}\int_{0}^{\pi}dk\,
\frac{\mathcal{K}_{e}(k)-\mathcal{K}_{e}^*(k)}{2}\mathcal{F}_{ec}(k).
\end{equation}
For real function $I(r)$ the standard definition (\ref{I-def})
remains valid, and new imaginary function $\Phi(r)$ has the form,
%
\begin{eqnarray}
&&\Phi(r)=\frac{1}{2\pi}\int_{-\pi}^{\pi}dk\,
\frac{\mathcal{K}_{e}(k)-\mathcal{K}_{e}^*(k)}{2\mathcal{K}_{ec}(k)}
\exp(-ikr)\,,\quad \quad \
\Re\Phi(r)=0\,;\label{ExpCos-Phi}\nonumber\\[6pt]
&&\frac{\mathcal{K}_{e}(k)-\mathcal{K}_{e}^*(k)}{2}
=i\,\frac{(\cos\omega-\cosh k_0\cos k)\sin\omega} {(\cosh
k_0\cos\omega-\cos k)^2+\sinh^2k_0\sin^2\omega}\,.
\label{ExpCos-Ke-Ke}
\end{eqnarray}

In order to find the values of constants $A$ and $B$, one should
write down two equations. The first one is nothing but the
normalization condition (\ref{A-eq}). The second equation can be
obtained by substitution of the solution (\ref{ExpCos-Fk-sol})
into the definition (\ref{ExpCos-B-def}). As a result, we come to
the system
%
\begin{eqnarray}
[I(0)+1]A+\Phi(0)B=2[I(0)-1]\,,\quad
\Phi(0)A-CB=2\Phi(0)\,,\label{ExpCos-Aeq}
\end{eqnarray}
where the real constant $C$ is given by the expressions,
\begin{eqnarray}\label{ExpCos-C-def}
C&=&2-\frac{1}{2\pi}\int_{0}^{\pi}dk\,
\frac{[\mathcal{K}_{e}(k)-\mathcal{K}_{e}^*(k)]^2}{2\mathcal{K}_{ec}(k)}
=1+\frac{1}{2\pi}\int_{-\pi}^{\pi}dk\,
\frac{\mathcal{K}_{e}(k)\mathcal{K}_{e}^*(k)}{\mathcal{K}_{ec}(k)}\,.
\end{eqnarray}
The solution of the system (\ref{ExpCos-Aeq}) is
%
\begin{eqnarray}
A=2\,\frac{I(0)-1+\Phi^2(0)/C}{I(0)+1+\Phi^2(0)/C}\,; \quad \quad
B=-\frac{4\Phi(0)/C}{I(0)+1+\Phi^2(0)/C}\,.\label{ExpCos-Asol}
\end{eqnarray}

Thus, the memory function $F(r)$ of the additive Markov chain with
exp-cosine correlations has the form,
\begin{eqnarray}\label{ExpCos-Fr}
F_{ec}(r)=\frac{2}{I(0)+1+\frac{\Phi^2(0)}{C}}
\Big\{[I(0)+\frac{\Phi^2(0)}{C}]\delta_{r,0}
-I(r)-\frac{\Phi(0)\Phi(r)}{C}\Big\}\,.
\end{eqnarray}
One can see that for $\omega=0$ the power spectrum is
$\mathcal{K}_{e}(k)=\mathcal{K}_{e}^*(k)=\mathcal{K}_{ec}(k)$, so
that we have $\Phi(r)=0$, and Eq.~(\ref{ExpCos-Fr}) coincides with
Eq.~(\ref{Exp-Fr-sol}).

\section{Power Spectrum with Simple Complex Poles}
\label{sec-CSP}

It is evident that the Fourier integrals (\ref{FTr-Kr}),
(\ref{FTr-Fr}) in the definitions of the binary correlator $K(r)$
and memory function $F(r)$ are completely determined by
singularities of corresponding Fourier transforms $\mathcal{K}(k)$
and $\mathcal{F}(k)$ in the complex plane of $k$. Taking into
account the periodicity conditions (\ref{K-par-per}) and
(\ref{F-par-per}), it is clear that the considered Fourier
integrals are determined by the singularities located within the
complex strip $(-\pi,\pi)$.

\begin{figure}
\begin{center}
\vspace{4.3cm}
\begin{center}
\includegraphics[width=9cm]{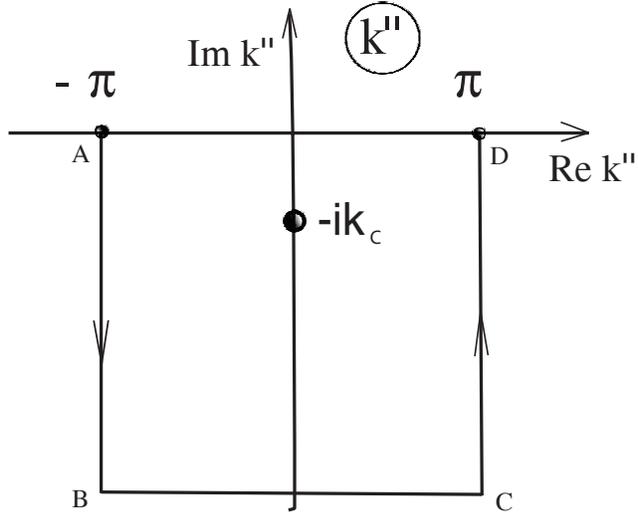}
\end{center}
\vspace{-4.3cm}
\end{center}
\caption{The contour of integration ABCD in the case of one
singularity in the strip.} \label{Fig}
\end{figure}

Let us consider the integral kernel $\Xi(k,k')$ represented by
Eq.(\ref{Ksi-2}). It is clear that $Q$-functions in the integrand
are analytical functions of the integration variable $k''$ in the
lower half-plane of $k''$. Indeed, they have periodically repeated
simple poles $k''=\pm|k|+i\epsilon$ and $k''=\pm|k'|+i\epsilon$
($\epsilon\to+0$) solely in the upper half-plane and converge to
zero when $k''\to-i\infty$, due to the expression,
\begin{eqnarray}\label{QQ-exp-small}
Q(k,k'')Q(k',k'')\approx 4\cos k\cos k'\exp(-2y),\nonumber\\[6pt]
\qquad\mbox{for}\,\,k''=-iy,\quad y\to\infty.
\end{eqnarray}
Therefore, the integral in Eq.~(\ref{Ksi-2}) is essentially
specified by the analytical properties of the power spectrum
$\mathcal{K}(k)$ in the lower half-plane. Let us evaluate the
integral along the closed contour $ABCDA$ with $A=-\pi-i0$,
$B=-\pi-i\infty$, $C=\pi-i\infty$ and $D=\pi-i0$ (see
Fig.~\ref{Fig}). The sum of integrals along the straight lines
$AB$ and $CD$ is equal to zero since the integrand is a periodic
function with the period $2\pi$. Then, the integral along the
infinitely far line $BC$ also vanishes because the product of two
last factors of the integrand vanishes exponentially fast at any
infinitely far point in the lower half-plane, see
Eq.~(\ref{QQ-exp-small}). Thus, the integral (\ref{Ksi-2}) taken
with the opposite sign, is completely determined by the
singularities of $\mathcal{K}(k)$ inside the considered contour
$ABCDA$, or the same, within the lower part of the complex strip
$(-\pi,\pi)$.

Let the power spectrum $\mathcal{K}(k)$ be analytical within the
considered strip. In this case $\mathcal{K}(k)$ is a constant and
due to the normalization condition (\ref{Kk-norm}), it has to be
one, $\mathcal{K}(k)=1$. From our analysis it follows,
$\Xi(k,k')=0$, that is in an accordance with the previously
obtained results for a white-noise disorder, see section
\ref{sec-WND}.

Now we are in a position to analyze the case when there is only
one singularity within the strip. Specifically, we consider a
simple pole at $k''=-ik_c$, assuming that the function
$\mathcal{K}(k)$ can by represented in the vicinity of this pole
by the form,
\begin{equation} \label{pole}
\mathcal{K}(k'')\simeq\frac{i}{k''+ik_c}.
\end{equation}
Here we take into account that the normalization condition
(\ref{Kk-norm}) requires the residue be equal $i$. In this case
one can get,
\begin{equation}\label{Ksi-exp-pole}
\Xi_e(k,k')=\frac{1}{2}Q(k,-ik_c)Q(k',-ik_c)\,.
\end{equation}
One can see that the above expression coincides with
Eq.~(\ref{Exp-Ksi}), thus reducing our problem to that considered
in Section VIII.

\section{Step-Wise Power Spectrum}
\label{sec-SWPS}

\subsection{General expressions}

As was mentioned in Section I, in the theory of correlated
disorder \cite{IzKr99,IzKrUll01,Lira,JPA}, one of important
problems is an additive binary Markov chain with a \emph{step-wise
power} (SWP) spectrum. The simplest case is given by the following
expressions,
%
\begin{eqnarray}
K(r)=h\,\delta_{r,0}+(1-h)\frac{\sin(k_cr)}{k_cr}\,,\label{SW-Kr}\\[6pt]
\mathcal{K}(k)=h+(1-h)\frac{\pi}{k_c}
\left[\Theta(k+k_c)-\Theta(k-k_c)\right]
=h+(1-h)\Delta_{k_c}(k)\,,\label{SW-Kk}\\[6pt]
\mathcal{K}^{-1}(k)=\frac{1}{h}
\Big[1-\frac{k_c(1-h)}{k_ch+\pi(1-h)}\,\Delta_{k_c}(k)\Big]\,,
\label{SW-Kk-1}\\[6pt]
0\leq h\leq1,\qquad0<k_c\leq\pi,\quad|k|\leq\pi.\nonumber
\end{eqnarray}
The restriction on the parameter $h$ is a consequence of a
non-negative nature of the power spectrum $\mathcal{K}(k)$.
Strictly speaking, the exact requirement is $0\leq
h\leq\pi/(\pi-k_c)$. In our analysis we assume $h$ be smaller than
one, in order to allow $k_c$ be very small.

In Eqs.~(\ref{SW-Kk}) and (\ref{SW-Kk-1}) the unit-pulse function
$\Delta_{k_c}(k)$ is defined by
%
\begin{eqnarray}
\Delta_{k_c}(k)=\frac{\pi}{k_c}\Theta(k_c-|k|),
\quad0<k_c\leq\pi,\quad|k|\leq\pi,\label{SW-Delta-def}\\[6pt]
\frac{1}{2\pi}\int_{-\pi}^{\pi}dk\Delta_{k_c}(k)=
\frac{1}{\pi}\int_{0}^{\pi}dk\Delta_{k_c}(k)=1,\label{SW-Delta-norm}\\[6pt]
\lim_{k_c\to0}\Delta_{k_c}(k)=\lim_{k_c\to0}\frac{\pi}{k_c}\Theta(k_c-|k|)
=2\pi\delta(k).\label{Theta-delta}
\end{eqnarray}
This definition and hence the power spectrum $\mathcal{K}(k)$,
together with its inverse quantity $\mathcal{K}^{-1}(k)$, contains
$\Theta(x)$ as the Heaviside unit-step function, $\Theta(x<0)=0$
and $\Theta(x>0)=1$. The factor $\pi/k_c$ in Eqs.~(\ref{SW-Kk})
and (\ref{SW-Delta-def}) provides the normalization condition
(\ref{Kk-norm}). As the majority of unit-pulse functions, our
$\Delta_{k_c}(k)$ is also a prelimit Dirac delta-function. This
fact is displayed by Eq.~(\ref{Theta-delta}). The Fourier
transforms for $\Delta_{k_c}(k)$ are
%
\begin{eqnarray}
\Delta_{k_c}(k)&=&\sum_{r=-\infty}^{\infty}\frac{\sin(k_cr)}{k_cr}\exp(ikr)
=1+2\sum_{r=1}^{\infty}\frac{\sin(k_cr)}{k_cr}\cos(kr);
\label{SW-Delta-sin}\\[6pt]
\frac{\sin(k_cr)}{k_cr}&=&\frac{1}{2\pi}\int_{-\pi}^{\pi}dk\Delta_{k_c}(k)\exp(-ikr)
=\frac{1}{\pi}\int_{0}^{\pi}dk\Delta_{k_c}(k)\cos(kr).\label{SW-sin-Delta}
\end{eqnarray}
Note that Eq.~(\ref{SW-Delta-sin}) for $\Delta_{k_c}(k)$ coincides
in form with the Fourier transforms (\ref{FTr-K-def}) for
$\mathcal{K}(k)$ and $K(r)$. This is not surprising since
\begin{eqnarray}\label{SW-KrKk-h0}
K(r)=\sin(k_cr)/(k_cr), \quad \quad
\mathcal{K}(k)=\Delta_{k_c}(k)\qquad \mbox{for}\quad h=0.
\end{eqnarray}
Therefore, one can expect that all formulas derived in this
section have to be reduced to the corresponding general ones in
the limit when $h=0$, with the change,
$\Delta_{k_c}(k)\to\mathcal{K}(k)$.

One can see that in the case when $k_c\to0$, we have the sequence
with long-range correlations, whereas if $k_c\to\pi$ the system is
a short-range correlated and at $k_c=\pi$ it reduces to the
white-noise (\ref{WN-Kr}). More generally, when $k_c$ diverges,
($k_c\to\infty$), the second term in Eq.~(\ref{SW-Kr}) decreases
with oscillations. At the same time, $K(r)$ gets the white-noise
form (\ref{WN-Kr}) every time when $k_c=n\pi$ ($n=1,2,3,\dots$).
We emphasize that Eqs.~(\ref{SW-Kk}) -- (\ref{SW-Kk-1}) are
applicable only when $k_c\leq\pi$. Note that the crossover to the
white-noise (\ref{WN-Kr}) can be also performed with $h\to1$.

In order to obtain the memory function $F(r)$ for the Markov chain
with the step-wise power spectrum (\ref{SW-Kr}), one should
firstly derive the kernel $\Xi(k,k')$ of the integral equation
(\ref{FTr-MF-eq}). For this, we substitute $K(r+r')$ or
$\mathcal{K}(k'')$ in the form (\ref{SW-Kr}) or (\ref{SW-Kk}) into
the definitions (\ref{Ksi-1}) or (\ref{Ksi-2}), respectively. The
first term with $h$ vanishes due to the condition (\ref{WN-Ksi}),
therefore, we have
%
\begin{eqnarray}
&&\Xi(k,k') =\frac{1-h}{2}\,4\sum_{r,r'=1}^{\infty}\cos(kr)
\frac{\sin[k_c(r+r')]}{k_c(r+r')}\cos(k'r')
\label{SW-Ksi-sum}\\[6pt]
&&=\frac{1-h}{2}\,
\frac{1}{2\pi}\int_{-\pi}^{\pi}dk''\Delta_{k_c}(k'')Q(k,k'')Q(k',k'')
\label{SW-Ksi-Delta}\\[6pt]
&&=\frac{1-h}{2}\,
\frac{1}{2k_c}\int_{-k_c}^{k_c}dk''Q(k,k'')Q(k',k'')\,.
\label{SW-Ksi-fin}
\end{eqnarray}
Now we employ Eq.~(\ref{QQ}) for the product of two Q-functions to
write explicitly some integrals. This yields,
\begin{equation}\label{SW-KsiI}
\Xi(k,k')=\frac{1-h}{2}\,\Big[1-\Delta_{k_c}(k)-\Delta_{k_c}(k')+
I_{sw}(k,k')\Big]\,.
\end{equation}
Here the new integral kernel reads,
%
\begin{eqnarray}
&&I_{sw}(k,k')=-\frac{1}{2\pi}\int_{-\pi}^{\pi}dk''\Delta_{k_c}(k'')
\times\frac{\sin^2k''}
{(\cos k-\cos k'')(\cos k'-\cos k'')}\label{Isw-Delta}\\[6pt]
&&=-\frac{1}{2k_c}\int_{-k_c}^{k_c}\frac{\sin^2k''dk''}
{(\cos k-\cos k'')(\cos k'-\cos k'')}\label{Isw-def}\\[6pt]
&&=1-\frac{1}{k_c}\,\frac{1}{\cos k-\cos k'}\Bigg[\sin k
\ln\Bigg|\frac{\sin[(k_c+k)/2]}{\sin[(k_c-k)/2]}\Bigg|
\nonumber\\[6pt]
&&-\sin k'
\ln\Bigg|\frac{\sin[(k_c+k')/2]}{\sin[(k_c-k')/2]}\Bigg|\Bigg].
\label{Isw1}
\end{eqnarray}
When $k_c=\pi$, we have $I_{sw}(k,k')=1$, and the kernel
$\Xi(k,k')$ vanishes. This fact is in an agreement with the
condition (\ref{WN-Ksi}). Actually, the same expression
(\ref{SW-KsiI}) for the kernel $\Xi(k,k')$ can be achieved from
Eq.~(\ref{Ksi-Ksi0}) by taking explicitly into account the
definition (\ref{Ksi0-def}) and Eq.~(\ref{Int4-pi}).

The kernel $I_{sw}(k,k')$ is a real, even and symmetrical function
of both variables $k$ and $k'$,
\begin{equation}\label{Isw-prop}
I_{sw}(-k,k')=I_{sw}(k,-k')=I_{sw}(k,k')=I_{sw}(k',k).
\end{equation}
In line with Eq.~(\ref{Int1}) taken at $k_c=\pi$ and
Eq.~(\ref{Int-Kkc}), this new kernel $I_{sw}(k,k')$ satisfies the
following integral properties,
%
\begin{eqnarray}
\frac{1}{\pi}\int_{0}^{\pi}dk\,I_{sw}(k,k')=\Delta_{k_c}(k')\,,
\label{Isw-int1}\\[6pt]
\frac{1}{\pi}\int_{0}^{\pi}dk'\,I_{sw}(k,k') =\Delta_{k_c}(k)\,.
\label{Isw-int2}
\end{eqnarray}
These relations together with Eq.~(\ref{SW-KsiI}) and
normalization condition (\ref{SW-Delta-norm}) provide the integral
properties (\ref{Int-Ksi1}), (\ref{Int-Ksi2}) of the old kernel
$\Xi(k,k')$.

Now we substitute Eqs.~(\ref{SW-Kk}) and (\ref{SW-KsiI}) into
Eq.(\ref{FTr-MF-eq}). One should take into account that two first
summands from $\Xi(k,k')$ do not contribute into the integral term
of Eq.~(\ref{FTr-MF-eq}), due to the normalization condition
(\ref{A-eq}). As a result, one can get the integral equation for
the memory function,
\begin{eqnarray}\label{SW-Fk-eq}
&&\frac{h}{1-h}\mathcal{F}(k)+\Delta_{k_c}(k)\mathcal{F}(k)
-\frac{1}{2\pi}\int_{0}^{\pi}dk'I_{sw}(k,k')\mathcal{F}(k')
\nonumber\\[6pt]
&&=\Delta_{k_c}(k)-1+\frac{1}{2}B_{sw}.
\end{eqnarray}
Here, instead of the old constant $A$, for the convenience we have
introduced the new one, $B_{sw}$, which can be associated with $A$
according to Eq.~(\ref{A-def}),
%
\begin{eqnarray}
&&A=(1-h)B_{sw},\label{A-Bsw}\\[6pt]
B_{sw}&=&\frac{1}{2\pi}\int_{-\pi}^{\pi}dk\,\Delta_{k_c}(k)\,\mathcal{F}(k)
=\frac{1}{2k_c}\int_{-k_c}^{k_c}dk\,\mathcal{F}(k).\label{Bsw-def}
\end{eqnarray}
Note that Eq.(\ref{SW-Fk-eq}) can be derived from
Eq.~(\ref{Fk-eq0}).

Finally, let us integrate Eq.~(\ref{SW-Fk-eq}) over the wave
number $k$ within the interval $(-\pi,\pi)$ or $(0,\pi)$. Then, on
its left-hand side the first term vanishes due to
Eq.~(\ref{A-eq}), second term turns into $B_{sw}$ in accordance
with its definition (\ref{Bsw-def}), while the last integral term
gives the value $B_{sw}/2$ because of the property defined by
Eqs.(\ref{Isw-int1}) and (\ref{Bsw-def}). On the right-hand side,
the first term compensates the unity. In such a manner, the
equation is reduced to the identity $B_{sw}/2\equiv B_{sw}/2$.

It should be, however, stressed that having Eq.(\ref{SW-Fk-eq}) in
its general form, it is still not clear how to derive the
solution.

\subsection{SWP Spectrum: Short-Range Correlations}
\label{subsec-SWPS-ShR}

A particular case is the SWP spectrum with short range
correlations, which corresponds to the values of $k_c$ close to
$\pi$. In this case Eqs.(\ref{SW-Kr}) and (\ref{SW-Kk}) result in
a particular case of Eq.~(\ref{SRCor-finFr}) with the integral
$I(r)$ that can be calculated in accordance with its definition
(\ref{I-def}). Taking into account Eq.~(\ref{SW-Kk-1}) one can
derive,
\begin{equation}\label{SW-I}
I(r)=\frac{1}{h}\,\left[\delta_{r,0}-
\frac{k_c(1-h)}{\pi-(\pi-k_c)h}\,\frac{\sin(k_cr)}{k_cr}\right].
\end{equation}
Therefore, the memory function has the form,
\begin{eqnarray}\label{SWShR-F}
F(r)&=&\frac{k_c(1-h)}{(\pi-k_c)(1-h)+k_ch}
\left[\frac{\sin(k_cr)}{k_cr}-\delta_{r,0}\right]\\[6pt]
&&\mbox{for}\qquad\pi-k_c\ll\pi.\nonumber
\end{eqnarray}

As a result, within the first approximation in $h\to0$ we get,
\begin{eqnarray} \label{SWShR-F-h0}
F(r)&=&\frac{k_c}{\pi-k_c}
\left[\frac{\sin(k_cr)}{k_cr}-\delta_{r,0}\right],
\end{eqnarray}
which is valid for $k_ch\ll\pi-k_c\ll\pi$.

\subsection{SWP Spectrum: Long-Range Correlations
(Zero Approximation)} \label{subsec-SWPS-LR0}

Now let us consider the more interesting case of the SWP spectrum
(\ref{SW-Kk}) with long-range correlations. In this case the
additive binary Markov chain has small wave number $k_c\ll1$.
Thus, within zero approximation in $k_c$ the exact expressions
(\ref{SW-Kk}) for the binary correlator $K(r)$ and its power
spectrum $\mathcal{K}(k)$ can be changed by the following
asymptotics,
%
\begin{eqnarray}
K(r)&=&h\,\delta_{r,0}+(1-h),\label{SWLR0-Kr}\\[6pt]
\mathcal{K}(k)&=&h+(1-h)2\pi\delta(k).\label{SWLR0-Kk}
\end{eqnarray}
with $|k|\leq\pi$. Note that these expressions are in an agreement
with Eqs.~(\ref{SW-Delta-def})-(\ref{Theta-delta}) and
(\ref{SW-Delta-sin})-(\ref{SW-sin-Delta}).

Now we substitute $K(r+r')$ or $\mathcal{K}(k'')$ in the form of
Eq.~(\ref{SWLR0-Kr}) or Eq.~(\ref{SWLR0-Kk}) into the definitions,
respectively, (\ref{Ksi-1}) or (\ref{Ksi-2}) for the kernel
$\Xi(k,k')$. Taking into account that the first term with $h$
vanishes due to the condition (\ref{WN-Ksi}), we get the result
%
\begin{eqnarray}
\Xi(k,k')=\frac{1-h}{2}\,\Big[2\sum_{r=1}^{\infty}\cos(kr)\Big]
\Big[2\sum_{r'=1}^{\infty}\cos(k'r')\Big]\label{SWLR0-Ksi-sum}\\[6pt]
=\frac{1-h}{2}\,Q(k,0)Q(k',0)\label{SWLR0-Ksi-Q}\\[6pt]
=\frac{1-h}{2}\,[1-2\pi\delta(k)][1-2\pi\delta(k')]\label{SWLR0-Ksi-del}\\[6pt]
=\frac{1-h}{2}\,\Big[1-2\pi\delta(k)-2\pi\delta(k')+
4\pi^2\delta(k)\delta(k')\Big]\,,\label{SWLR0-Ksi-Df}
\end{eqnarray}
that should be compared with the exact
expressions~(\ref{SW-Ksi-sum})-(\ref{SW-Ksi-fin}) and
(\ref{SW-KsiI}). In order to pass from Eqs.~(\ref{SWLR0-Ksi-sum})-
(\ref{SWLR0-Ksi-Q}) to Eq.~(\ref{SWLR0-Ksi-del}) and then to
Eq.~(\ref{SWLR0-Ksi-Df}), one should recognize that from the
definition (\ref{Q-def1}) for the $Q$-function, it follows,
\begin{equation}\label{Q-k0}
Q(k,0)=-2\sum_{r=1}^{\infty}\cos(kr)=1-2\pi\delta(k).
\end{equation}
Note that the expressions
(\ref{SWLR0-Ksi-sum})-(\ref{SWLR0-Ksi-Df}) and (\ref{Q-k0}) fulfil
the general relations (\ref{Int-Ksi1})-(\ref{Int-Ksi2}).

The integral equation (\ref{FTr-MF-eq}) with
Eq.(\ref{SWLR0-Ksi-Df}) taken as the kernel $\Xi(k,k')$, gives
rise to the equation,
\begin{eqnarray}\label{SWLR0-Fk-eq}
\frac{h}{1-h}\,\mathcal{F}(k)+2\pi\delta(k)\mathcal{F}(k)-
\pi\delta(k)\mathcal{F}(0)
=2\pi\delta(k)-1+\frac{1}{2}\,\mathcal{F}(0).
\end{eqnarray}
Here we take into account that
\begin{equation}\label{Bsw-LR0}
B_{sw}=\mathcal{F}(0)\qquad\mbox{for}\quad k_c\to0.
\end{equation}
All terms in the equation (\ref{SWLR0-Fk-eq}) correspond to those
in the general equation (\ref{SW-Fk-eq}). Thus, the solution for
the memory function yields
%
\begin{eqnarray}
\mathcal{F}(k)&=&\frac{1-h}{h}\,
\Big[1-\frac{1}{2}\,\mathcal{F}(0)\Big]
[2\pi\delta(k)-1]\,,\label{SWLR0-Fk}\\[6pt]
F(r)&=&\frac{1-h}{h}\,\Big[1-\frac{1}{2}\,\mathcal{F}(0)\Big]
[1-\delta_{r,0}]\,,\,\, k_c\to0.\label{SWLR0-Fr}
\end{eqnarray}
In principle, these relations would give a complete solution of
the problem. However, a problem remains how to find the constant
$B_{sw}=\mathcal{F}(0)$. Indeed, the solution
(\ref{SWLR0-Fk})-(\ref{SWLR0-Fr}) automatically satisfies to the
normalization condition (\ref{A-eq}). On the other hand, the
relation (\ref{SWLR0-Fk}) violates if one takes $k=0$.

\subsection{SWP Spectrum: Long-Range Correlations (Another Approach)}
\label{subsec-SWPS-LRM}

Our failure to derive the correct solution in previous subsection
is evidently related to the fact that at the beginning we have
substituted the rigorous expressions (\ref{SW-Kr})-(\ref{SW-Kk})
by Eqs.~(\ref{SWLR0-Kr})-(\ref{SWLR0-Kk}). Strictly speaking, the
latter are always invalid since the smaller $k_c$ the larger
indices $r\sim k_c^{-1}$ play an important role in
Eq.~(\ref{SW-Kr}) and in the sum of Eq.~(\ref{SW-Ksi-sum}). This
results in the value of $\Delta_{k_c}(k)$, not in $2\pi\delta(k)$,
in the expression (\ref{SW-Kk}) for the power spectrum and
(\ref{SW-Ksi-Delta}) for the integral kernel, see also
Eqs.~(\ref{SW-Delta-sin})-(\ref{SW-sin-Delta}).

In order to improve the approach formulated above, let us replace,
respectively, the Dirac delta-functions $2\pi\delta(k)$ and
$2\pi\delta(k')$ with prelimit ones $\Delta_{k_c}(k)$ and
$\Delta_{k_c}(k')$ in asymptotical Eqs.~(\ref{SWLR0-Ksi-del}) and
(\ref{SWLR0-Ksi-Df}). With this, the expression for the kernel
turns out to have the form,
%
\begin{eqnarray}
\Xi(k,k')&=&\frac{1-h}{2}\,4\sum_{r,r'=1}^{\infty}\cos(kr)
\frac{\sin(k_cr)}{k_cr}\nonumber\\[6pt]
&&\times\frac{\sin(k_cr')}{k_cr'}\cos(k'r')
\label{SWLRM-Ksi-sum}\\[6pt]
&=&\frac{1-h}{2}\,[1-\Delta_{k_c}(k)][1-\Delta_{k_c}(k')]
\label{SWLRM-Ksi-Delta}\,.
\end{eqnarray}
Let us compare the first expression (\ref{SWLRM-Ksi-sum}) with
\emph{exact} Eq.~(\ref{SW-Ksi-sum}) and zero-asymptotical
Eq.~(\ref{SWLR0-Ksi-sum}). In the \emph{exact} expression
(\ref{SW-KsiI}) for the kernel $\Xi(k,k')$ in the case of
long-range correlations, $k_c \ll 1$, we admit that the following
replacement
\begin{equation}\label{Isw-LRM}
I_{sw}(k,k')\to\Delta_{k_c}(k)\Delta_{k_c}(k')
\end{equation}
is correct. Note that such a model conserves all general
properties of the function $I_{sw}(k,k')$ described above after
Eqs.~(\ref{Isw-Delta})-(\ref{Isw-def}), and as a consequence,
provides all properties of the kernel $\Xi(k,k')$, as well as the
exact equations (\ref{FTr-MF-eq}) and (\ref{SW-Fk-eq}).

The assumption (\ref{Isw-LRM}) allows one straightforwardly to
obtain the solution of the equation (\ref{SW-Fk-eq}),
\begin{eqnarray}\label{SWLRM-Fk1}
&&\frac{2h}{1-h}\,\mathcal{F}(k)
=(2+B_{sw})\Big[1+\frac{1-h}{h}\,\Delta_{k_c}(k)\Big]^{-1}\Delta_{k_c}(k)
\nonumber\\[6pt]
&&-(2-B_{sw})\Big[1+\frac{1-h}{h}\,\Delta_{k_c}(k)\Big]^{-1}.
\end{eqnarray}
Now one should take into account the equalities
\begin{eqnarray}\label{SWLR-Inv}
\Big[1+\frac{1-h}{h}\,\Delta_{k_c}(k)\Big]^{-1}
=1-\frac{k_c(1-h)}{k_ch+\pi(1-h)}\,\Delta_{k_c}(k)\,,
\label{SWLR-Inv1}\\[6pt]
\Big[1+\frac{1-h}{h}\,\Delta_{k_c}(k)\Big]^{-1}\Delta_{k_c}(k)
\qquad\qquad\nonumber\\[6pt]
=\frac{k_ch}{k_ch+\pi(1-h)}\,\Delta_{k_c}(k)\,,
\,\,\Delta_{k_c}^2(k)=\frac{\pi}{k_c}\Delta_{k_c}(k)\,.
\label{SWLR-Inv2}
\end{eqnarray}
Then the solution gets the form
\begin{eqnarray}\label{SWLRM-Fk2}
&&\frac{2h}{1-h}\,\mathcal{F}(k) =-(2-B_{sw})\nonumber\\[6pt]
&&+\frac{4k_ch-(2-B_{sw})[k_ch-k_c(1-h)]}{k_ch+\pi(1-h)}
\,\Delta_{k_c}(k).
\end{eqnarray}
From the normalization condition (\ref{A-eq}) one can obtain
\begin{equation}\label{SWLRM-A}
2-B_{sw}=\frac{4k_ch}{(\pi-k_c)(1-h)+2k_ch}\,.
\end{equation}

As a result, in the case of long-range correlations we have
%
\begin{eqnarray}
\mathcal{F}(k)=\frac{2k_c(1-h)}{(\pi-k_c)(1-h)+2k_ch}
[\Delta_{k_c}(k)-1],\label{SWLRM-Fk}\\[6pt]
F(r)=\frac{2k_c(1-h)}{(\pi-k_c)(1-h)+2k_ch}
\left[\frac{\sin(k_cr)}{k_cr}-\delta_{r,0}\right]\label{SWLRM-Fr}
\end{eqnarray}
Evidently, this expression can be simplified by making use of a
smallness of $k_c$. Thus, the final correct expression gets the
form,
%
\begin{eqnarray}
\mathcal{F}(k)=\frac{2k_c}{\pi}
[\Delta_{k_c}(k)-1]\,,\label{SWLRM-Fkh0}\\[6pt]
F(r)=\frac{2k_c}{\pi}
\left[\frac{\sin(k_cr)}{k_cr}-\delta_{r,0}\right].\label{SWLRM-Frh0}
\end{eqnarray}
Note that the obtained result
(\ref{SWLRM-Fkh0})-(\ref{SWLRM-Frh0}) is independent of $h$, hence
it is also valid for $h=0$.

\section{Concluding Remarks}

In this paper we suggest a method of construction of a binary
Markov chain with a prescribed pair correlator. It is based on the
integral relation between the pair correlator and the so-called
memory function. The knowledge of the latter allows one to create
binary sequences that have given short or long-range pair
correlations. Our interest in this method is mainly due to its
application to the problem of selective transport in
one-dimensional disordered systems. As is known, the theory of
quantum transport through finite samples described by a
white-noise potential is fully developed. The key ingredient is
the well known Anderson localization, that states that all
eigenstates in infinite samples are exponentially localized
regardless of the strength of disorder. The knowledge of the
localization length of these eigenstates allows for a complete
statistical description of all transport properties of finite
samples, due to the single-parameter scaling conjecture.

In contrast with the white noise disorder, the role of long-range
correlations in random potentials is far from being properly
understood. Recent results reveal quite unexpected properties of
the transport in systems with such potentials. In particular, in
Refs.\cite{IzKr99,IzKrUll01,Lira,JPA,KIKS00,KIKS02,IzMak01,IzMak02PIERS,IzMak03}
it was shown that specific long-range correlations give rise in a
very sharp change of the transmission coefficient, when the energy
of an incident wave crosses some value $E_c$. This effect is
similar to the mobility edge effect in three-dimensional solid
state models, according to which on one side of $E_c$ all
eigenstates are localized, in contrast with other side where the
eigenstates are extended. It is assumed that analogous effects in
one-dimensional geometry may find various applications, such as a
construction of materials with a highly selective transport or
explanation of a selective conductivity in DNA molecules.

However, as was recently understood, an extension of this theory
to binary sequences suffers from unexpected problems. We hope that
our study of the integral equation (\ref{MF-BC-eqini}) relating
the memory function and pair correlator can shed light on the
problem of binary correlated disorder. Since there is no general
theory for solving integral equations, we have decided to start
with particular cases allowing for analytical solutions. In this
line, apart from the study of general properties of the main
equation (\ref{MF-BC-eqini}), we have solved relatively simple,
however, physically interesting cases of short-range,
exponentially decaying, and exponential-cosine correlations. In
all these cases we did not met serious mathematical problems. On
the other hand, a more general problem of the step-function
(\ref{SW-Kk}) that is of specific interest in view of the mobility
edge transition, we were able to solve this problem for the cases
when the mobility edge is close to either one or other energy
band. We would like to stress that these two cases are also
interesting from the view point of different applications, giving
close analytical expressions for the pair correlators. We have
found, however, that it is not clear how to obtain the general
solution of the problem. It is still not clear whether the
mathematical difficulties of solving of the integral equation
(\ref{MF-BC-eqini}) are principal, or they can be solved in a
different approach. Note, that here we discuss the analytical
results only, leaving aside a possibility to solve the problem
numerically. We plan to do this in our next studies.

To conclude with, we would like to note that the problem of
construction of sequences with a power-law decay of the pair
correlator, is quite specific. In particular, one should take into
account strong restrictions imposed by a rigorous theorem stating
that the Fourier transform of any pair correlator is a
non-negative function (this corresponds to the fact that the
Lyapunov exponent of the considered processes can not be
negative). This results in a quite unexpected conclusions. For
example, there are no sequences giving rise to a monotonic power
decay for the pair correlator, such as $ K(r) = C/r^{\gamma}$,
where $C$ and $\gamma$ are some constants. In other words, one
should always take into account the oscillations of a pair
correlator, such as $\sin(k_c r)$. These oscillations play an
important role providing the non-negative values of the Lyapunov
exponents (see, for example, Ref.\cite{IzKr99}). In the latter
case their influence on physical observables may be neglected.
Therefore, it is not clear whether one can speak about "generic
cases" of the power decaying correlations.


\section{Acknowledgements}

This research was partially supported by the CONACYT (M\'exico)
grant No~43730, and by the VIEP-BUAP (M\'exico) under the grant
5/G/ING/06.


\end{document}